\documentclass[prd,aps,floats,preprint,superscriptaddress,nofootinbib,tightenlines]{revtex4}
\usepackage{epsfig}
\usepackage{amsmath}
\def\euv{ \varepsilon_{\text{UV}} }
\def\eir{ \varepsilon_{\text{IR}} }

\def\e{\varepsilon}

\def\Dsl{\hbox{/\kern-.6000em D}} 

\def\dsl{\,\raise.15ex\hbox{/}\mkern-13.5mu D}

\def\ltap{\ \raise.3ex\hbox{$<$\kern-.75em\lower1ex\hbox{$\sim$}}\ }
\def\gtap{\ \raise.3ex\hbox{$>$\kern-.75em\lower1ex\hbox{$\sim$}}\ }
\def\OMIT#1{}

\def\OMIT#1{}

\newcommand{\nn}{\nonumber}

\newcommand{\bea}{\begin{eqnarray}}
\newcommand{\eea}{\end{eqnarray}}
\newcommand{\nb}{\bar n}

\newcommand{\pp}{\prime \prime}

\begin{document}
\title{Demonstration of the Equivalence of Soft and Zero-Bin Subtractions}

\author{Ahmad Idilbi}
\email{idilbi@phy.duke.edu}
\affiliation{Department of Physics, Duke University, Durham NC 27708, USA}

\author{Thomas Mehen}
\email{mehen@phy.duke.edu}
\affiliation{Department of Physics, Duke University, Durham NC 27708, USA}
\affiliation{Jefferson Laboratory, 12000 Jefferson Ave., Newport News VA 23606, USA}
\date{\today}
\vspace{0.5in}
\begin{abstract}

Calculations of collinear correlation functions in perturbative QCD and Soft-Collinear Effective Theory (SCET)
require a prescription for subtracting  soft or zero-bin contributions in order to avoid double counting the 
contributions from soft modes. At leading order in $\lambda$, where $\lambda$ is the SCET expansion  parameter, the
zero-bin subtractions have been argued to be equivalent to convolution with soft Wilson lines. We give a proof of the
factorization of naive collinear Wilson lines that is crucial for the derivation of the equivalence. We then check
the equivalence by computing the non-Abelian two-loop mixed collinear-soft contribution to the jet function in the
quark form factor.  These results demonstrate the equivalence, which can be used to give a
nonperturbative definition of the zero-bin subtraction at lowest order in $\lambda$.

\end{abstract}

\maketitle

In perturbative QCD (pQCD) factorization theorems~\cite{Collins:1989gx}, a cross section is expressed as a convolution
of several distinct functions, each of which captures physics at a given scale.
Suppose a process contains a hard scattering characterized by a scale $Q$
which is much greater than $\Lambda_{\rm QCD}$. The factorization
theorem typically contains perturbatively calculable hard coefficients, which capture physics at the scale $Q$,
 jet or collinear functions, which describe the propagation of particles in the initial or final state
with energies of order $Q$ but whose invariant mass is typically $O(\Lambda_{\rm QCD})$
or $O(\sqrt{\Lambda_{\rm QCD} Q})$, and soft functions which describe low energy quanta emitted
in the process.

Soft-Collinear Effective Theory (SCET)~\cite{Bauer:2000ew,Bauer:2000yr,Bauer:2001yt}
is an effective theory that can be used to derive
factorization theorems in QCD. In this approach to QCD factorization, QCD is matched onto
an effective theory that contains collinear and soft degrees of freedom, whose momentum
components in light-cone coordinates scale as 
\bea \label{scaling}
{\rm collinear:} \qquad (\nb\cdot p,  n\cdot p ,p^\perp) &\sim& Q(1,\lambda^2, \lambda) \, \nn\\
{\rm soft:}  \qquad (\nb\cdot p,  n\cdot p ,p^\perp) &\sim& Q(\lambda^2,\lambda^2, \lambda^2) \, , 
\eea
where $\lambda \sim \sqrt{\Lambda_{\rm QCD}/Q}$. Here the light-like vectors are 
$\nb^\mu = (1,0,0,-1)$ and $n^\mu =(1,0,0,1)$. This is the  power counting of  SCET$_{\rm I}$ which
is relevant for inclusive processes.  For exclusive processes, a different power counting is 
needed, and the appropriate effective theory is SCET$_{\rm II}$~\cite{Bauer:2002aj}. 
Below when we refer to SCET, it is implied that we are discussing SCET$_{\rm I}$.
Factorization theorems in this approach are obtained by matching QCD onto SCET at 
the hard scale, $Q$, then decoupling the soft and collinear degrees of freedom in the effective theory by a field 
redefinition~\cite{Bauer:2001yt}. The matching coefficients in the effective field theory are the hard coefficients of the traditional
pQCD factorization theorems. Correlation functions of the collinear and soft fields correspond to the jet functions
and soft functions, respectively. SCET is formulated as a systematic expansion in $\lambda$
and therefore provides a framework for analyzing power corrections to leading twist pQCD factorization 
theorems.

In the evaluation of loop or phase space integrals that arise in the perturbative calculation of
soft or jet functions, one could impose cutoffs to enforce the constraints of Eq.~(\ref{scaling}) so
that no virtual soft modes contribute to the calculation of a jet function or vice versa. In practice
this would make integrals horribly complicated and one almost always integrates over all momentum 
space~\cite{Beneke:1997zp}. This then raises the important question of avoiding double counting of soft 
contributions in both collinear and soft functions. This problem arises in both pQCD and SCET 
formulations of the problem. The traditional approach in pQCD has been to argue that jet functions 
need to be convolved with soft Wilson lines~\cite{Akhoury:1998gs,Ji:2004wu,Collins:1989bt,Collins:1999dz,
Chen:2006vd}, 
while in SCET this is implemented by ``zero-bin'' subtractions~\cite{Manohar:2006nz}.

An argument for the  equivalence (to lowest order in $\lambda$) of the two formalisms  was first given in Ref.~\cite{Lee:2006nr}.
The essence of the argument of Ref.~\cite{Lee:2006nr} is that  the zero-bin mode of the naive collinear field can be decoupled from
the purely collinear field by a field redefinition similar to that used to decouple soft modes from collinear modes in SCET. 
Performing this field redefinition on the naive collinear matrix element, one finds that the naive jet  function factorizes into a
purely collinear function convolved with a vacuum matrix element of a zero-bin or soft Wilson line. In a previous
paper~\cite{Idilbi:2007ff}, we studied the equivalence of soft and zero-bin subtractions in the quark form factor and in deeply
inelastic scattering (DIS) near  $x \to 1$, where $x$ is the Bjorken variable. We emphasized the importance of using an infrared
(IR) regulator such as dimensional regularization (DR) which does not take external particles off-shell and therefore spoil the field
redefinition that relates the naive and purely collinear functions. We checked the equivalence at one-loop for  both the quark form
factor and DIS as $x \to 1$. We also verified  the equivalence of soft and zero-bin  subtractions for the two-loop Abelian diagrams
contributing to the jet function appearing in the quark form factor.

In this paper we complete the analysis initiated in Ref.~\cite{Idilbi:2007ff}. Our first objective 
is to complete the argument for the equivalence of soft and zero-bin subtractions
presented in Ref.~\cite{Lee:2006nr}. A crucial step in the argument is that the Wilson line of the naive 
collinear gluons factorizes into the product of a Wilson line constructed from the collinear zero-bin
and a purely collinear Wilson line. This factorization was assumed in the derivation of Ref.~\cite{Lee:2006nr}
but not derived. In this paper, we provide a derivation of this factorization, valid up to corrections of 
order $\lambda^2$. Our second objective is to complete the 
analysis of zero-bin subtractions in the jet function of the quark form factor~\cite{Idilbi:2007ff}.
In particular, we verify  the equivalence of soft and zero-bin subtractions to two-loops in the non-Abelian theory.

The results of this paper demonstrate the equivalence of the soft Wilson line subtraction
of pQCD and the zero-bin subtractions in SCET. It is satisfying to understand the relationship between
the two approaches to  avoiding double counting. The equivalence of soft and zero-bin subtractions
may simplify the calculation of higher order loop diagrams in SCET.
In addition, the equivalence
provides a nonperturbative operator definition of the SCET zero-bin subtraction, which was defined 
diagrammatically in Ref.~\cite{Manohar:2006nz}. 

The paper is organized as follows. In the next section we review the argument of Ref.~\cite{Lee:2006nr}
for the equivalence and provide the proof required for the factorization of the naive collinear
Wilson line. In Section III, we complete the analysis of the two-loop zero-bin contribution to the 
jet function in the factorization theorem for the quark form factor in non-Abelian gauge theory. 
The equivalence of the zero-bin subtraction and dividing by the soft Wilson lines requires that 
mixed collinear and soft zero-bin contributions from certain Feynman diagrams proportional to 
$C_F C_A$ must add up to zero. This cancellation is verified in this section.
Section IV contains our conclusions. In the Appendix we discuss some subtleties 
in evaluating two-loop zero-bin subtractions that arise in the $L_2$ limit~\cite{Idilbi:2007ff}.

\section{Factorization of Collinear Wilson Lines}

In this section we briefly review the argument of Ref.~\cite{Lee:2006nr} and supply a proof of the 
factorization of the naive collinear Wilson line. Consider the naive collinear matrix element,
\bea\label{ncm}
\langle X_n| \bar \xi_n^{\pp} W_n^{\pp}|0\rangle \,.
\eea
Here $\xi_n^{\pp}$ is the naive collinear field and $W_n^{\pp}$ is the naive collinear Wilson line.
By naive we mean that the zero-bin mode has not been removed, either from the definitions of the 
fields or the SCET Lagrangian.  The final state, $\langle X_n|$,  contains collinear 
quanta. For the (unphysical) quark form factor, $\langle X_n |$ contains a single collinear quark, while 
for a physical quantity there  will be a sum over infinitely many particles, weighted by a shape variable.
We will not specify any properties of $\langle X_n|$ since they are not required.
Lee and Sterman~\cite{Lee:2006nr} argue that the zero-bin mode couples to the purely collinear modes in the same way as a soft mode, 
so the zero-bin modes can be decoupled from purely collinear modes by a similar field redefinition. Let 
\bea
U_n(x) = P \exp\left[i g \int_{0}^{\infty} ds\,  n\cdot A_{n,0}^{\pp}(ns +x)\right] \, ,
\eea
be the Wilson line constructed from the zero-bin mode of the collinear gauge field.
The zero-bin mode  is decoupled from other collinear modes by the field redefinition  
\bea
\xi_n^{\pp} = U_n^\dagger \xi_n^\prime, \qquad A_n^{\pp} = U_n^\dagger A_n^\prime U_n,
\qquad W_n^{\pp} = U_n^\dagger \tilde W_n^\prime U_n \,.
\eea
Though the $n\cdot A^{\pp}_{n,0}$ collinear zero-bin has been decoupled,
$\tilde W^\prime_n$ still contains zero-bin modes of the field $\nb \cdot A^\prime_n$.
To make the dependence of these modes explicit in what follows, we define 
$\nb \cdot A_{n,0}$ (without a prime) to be the zero-bin mode of $\nb \cdot A^\prime_n$ and from now on use the 
notation $\nb \cdot A^\prime_n$ to refer only to the purely collinear contribution,
i.e., $\nb \cdot A_n^\prime = \sum_{q \neq 0}\nb \cdot A_{n,q}$. The explicit 
dependence on $\nb \cdot A_{n,0}$ can be extracted because of the following property of 
$\tilde W_n^\prime$,
\bea\label{fact}
\tilde W_n^{\prime} (x)= W_n^\prime(x) \Omega_n(x) \, ,
\eea
which, as we will see below, holds up to corrections of $O(\lambda^2)$.
Here, $W_n^\prime$ is the purely collinear Wilson line which does not contain zero-bin modes,
and $\Omega_n$ is another zero-bin Wilson line defined by
\bea
\Omega_n(x) = P \exp\left[i g \int_{-\infty}^0 ds\, \nb \cdot A_{n,0}(\nb s +x)\right] \, .
\eea
Using Eq.~(\ref{fact}) one finds that the naive collinear matrix of Eq.~(\ref{ncm}) factorizes as 
\bea
\langle X_n| \bar \xi_n^{\pp} W_n^{\pp}|0\rangle 
=\langle X^\prime_n| \bar \xi_n^\prime W_n^\prime|0\rangle \, \langle X^0_n| \Omega_n U_n|0\rangle \, .
\eea
We have factorized the final state, $\langle X_n |$, into the product of a state that contains 
purely collinear degrees of freedom, $\langle X^\prime_n|$, and a state containing zero-bin
modes only, $\langle X^0_n|$. We see that the naive collinear matrix element factors into the product of 
a purely collinear matrix element and a matrix element of zero-bin Wilson lines.

In Ref.~\cite{Lee:2006nr} the factorization in Eq.~(\ref{fact}) is simply assumed, no argument for its validity
is given. Here we fill this gap. The Wilson lines $\tilde W_n^\prime$ and $\Omega_n$ obey the following 
differential equations:
\bea\label{ihde}
(i \nb \cdot \partial + g \,\nb \cdot A_{n,0} + g \, \nb \cdot A_n^\prime)\tilde W_n^\prime &=& 0  \\
(i \nb \cdot \partial + g \, \nb \cdot A_{n,0})\Omega_n &=& 0
\eea
Note that Eq.~(\ref{ihde}) is not homogeneous in the power counting parameter $\lambda$.
The field $\nb \cdot A_{n,0}$ is $O(\lambda^2)$, $\nb \cdot A^\prime_n$ is $O(1)$, and 
$i \nb \cdot \partial$ has no definite scaling with $\lambda$ since it can act on either
the $\nb \cdot A_{n,0}$ or $\nb \cdot A^\prime_n$ present in $\tilde W_n^\prime$. The factorization of Eq.~(\ref{fact}) 
separates $\tilde W_n^\prime$ into Wilson lines that are solutions to first order
equations that are homogeneous in $\lambda$. To obtain this factorization, we define 
the function $W_n^t$ by
\bea\label{def2}
\tilde W_n^{\prime} (x)= \Omega_n(x) W_n^t(x) \, .
\eea
Applying the chain rule to $W_n^t(x) = \Omega_n^\dagger(x) \tilde W_n^\prime(x)$,
it is straightforward to show that 
\bea
(i \nb \cdot \partial + g\, \Omega_n^\dagger \nb \cdot A_n^\prime \Omega_n) W_n^t &=& 0 \, ,
\eea
so that 
\bea\label{wnt}
W_n^t(x) &=& P\exp\left[i g \int_{-\infty}^0 ds\, \Omega_n^\dagger(\nb s +x) \nb \cdot A^\prime_n(\nb s +x) \Omega_n(\nb s +x)\right] \nn \\
&=& P\exp\left[i g \int_{-\infty}^0 ds\, \Omega_n^\dagger(x) \nb \cdot A^\prime_n(\nb s +x) \Omega_n(x)\right] + O(\lambda^2) \nn \\
&=& \Omega_n^\dagger(x)  P\exp\left[i g \int_{-\infty}^0 ds\, \nb \cdot A^\prime_n (\nb s +x)\right] \Omega_n(x) + O(\lambda^2) \nn \\
&=& \Omega_n^\dagger(x)  \, W_n^\prime(x) \, \Omega_n(x)\, + O(\lambda^2) .
\eea
Dropping terms supressed by $\lambda^2$ and plugging the result back into Eq.~(\ref{def2}), we obtain the desired factorization formula 
for $\tilde W_n^\prime$. The second line of Eq.~(\ref{wnt}) follows as a consequence of 
power counting, since 
\bea\label{approx}
\Omega_n(\nb s +x) = \exp(s \nb\cdot \partial)\Omega_n(x) =\Omega_n(x) +O(\lambda^2) \, .
\eea
If we try to write $\nb \cdot A^\prime_n(\nb s +x )$ as a power series in $s$, $\nb \cdot A^\prime_n(\nb s +x )
= \exp(s \nb\cdot \partial)\nb \cdot A^\prime_n(x)$, all terms in the exponential are $O(1)$ since 
$\nb  \cdot A^\prime_n$ is purely collinear. The approximation of Eq.~(\ref{approx}) is equivalent
to dropping $O(\lambda^2)$ zero-bin momenta relative to $O(1)$ purely collinear label momenta
in the evaluation of the Wilson line in momentum space.

A physical jet function is not of the form of Eq.~(\ref{ncm}), but rather the square of such 
a matrix element with a sum over final states. For example, the jet function in a factorization theorem 
for an event shape cross section takes the form~\cite{Lee:2006nr},
\bea\label{jet}
J^\prime_n(e) =\sum_{X_n} |\langle X_n|\bar \xi^\prime W_n^\prime |0\rangle|^2 \delta(e -e(X_n)) \, ,
\eea
where $e$ is the event shape variable and the delta-function ensures that only final states with
$e$ are summed over. Applying the factorization of the collinear matrix element in 
Eq.~(\ref{ncm}) to jet functions, one finds that the naively evaluated jet function can be expressed
as a convolution of the purely collinear jet function in Eq.~(\ref{jet}) and an eikonal 
jet function, which is defined in terms of matrix elements of the product of zero-bin
Wilson lines, $\Omega_n U_n$~\cite{Lee:2006nr}. This convolution can be rendered a simple
product in moment space, so one obtains 
\bea\label{moment}
\tilde J^\prime_n(N) = \frac{\tilde J^{\pp}_n(N)}{\tilde J^{\rm eik}_n(N)}\, ,
\eea
where $\tilde J^\prime_n(N)$, $\tilde J^{\pp}_n(N)$, and $\tilde J^{\rm eik}_n(N)$ are the $N$th-moments of the purely collinear,
naive collinear, and eikonal (i.e. soft)  jet functions. If the jet function of interest is sufficiently inclusive  that the event
shape only depends on the total momentum of the final state  collinear particles (e.g., the factorization theorem for DIS as $x \to
1$), then one can  use translation invariance and completeness to write the jet function as the Fourier transform of a $T$-ordered
product of the  operators $\bar \xi_n^\prime W_n^\prime $ and  $W_n^{\prime \,\dagger } \xi^\prime_n$. The naive and purely collinear
jet functions in this case are related by an equation analogous to Eq.~(\ref{moment}), where all jet functions are now defined by
the corresponding $T$-ordered products. For sufficiently inclusive processes the analog of ${\tilde J}_n^{\rm eik}(N)$ is the
well-known soft function. Therefore the work of Ref.~\cite{Lee:2006nr} combined with the results of this section 
provides an effective field theory demonstration of how dividing by the soft function eliminates double counting in pQCD factorization 
theorems for inclusive processes.

\section{Two-Loop Zero-Bin Subtractions for the Jet Function}

Applying the arguments of the last section to the incoming jet function that appears in the SCET factorization theorem
for the quark form factor~\cite{Idilbi:2007ff}, one finds that
\bea
\label{exp}
  \langle 0| W_n^{\prime \, \dagger}  \xi_n^\prime |q(p)\rangle
  =\frac{\langle 0| W_n^{\pp \, \dagger}\xi_n^{\pp}|q(p)\rangle}{\langle 0 |U_n^\dagger \Omega_n^\dagger  |0\rangle}\, .
\eea 
Note that, unlike Section I, we are considering a jet function with a particle in the initial
rather than final state. This quantity is clearly unphysical, however, it can be used to test the equivalence of soft
and zero-bin subtractions since the argument for the  factorization of the naively collinear matrix element
into purely collinear and zero-bin matrix elements is independent of the initial and final states.
The zero-bin subtractions needed for the purely collinear matrix element, $\langle 0| W_n^{\prime \, \dagger}  \xi_n^\prime |q(p)\rangle$,
are the same zero-bin subtractions needed for virtual contributions to a physical jet function.

The remainder of this paper is devoted to completing the two-loop check of Eq.~(\ref{exp}), initiated
in Ref.~\cite{Idilbi:2007ff}. Specifically, we wish to calculate the two-loop zero-bin subtractions
for the left hand side of Eq.~(\ref{exp}) and verify that they are reproduced by the right hand side.
To $O(\alpha_s^2)$ the matrix elements on the right hand side of Eq.~(\ref{exp}) can be parameterized as:
\bea\label{tlexp}
\frac{\langle 0| W_n^{\pp \, \dagger}\xi_n^{\pp}|q(p)\rangle}{\langle 0 |U_n^\dagger \Omega_n^\dagger|0\rangle}
&=&\frac{1+ \alpha_s\, C_F \,I^{(1)}_n +  \alpha_s^2 \,C_F^2 I^{(2)}_{n,C_F^2}+ \alpha_s^2 \, C_F C_A I_{n,C_F C_A}^{(2)} +O(\alpha_s^3)}
{1+ \alpha_s \,C_F I^{(1)}_s+\frac{1}{2} \alpha_s^2 \,  C_F^2 [I^{(1)}_s]^2 + \alpha_s^2 \, C_F C_A I_{s, C_F C_A}^{(2)}
 +O(\alpha_s^3)} \nn \\
&=&1 + \alpha_s \,C_F \,(I^{(1)}_n -I_s^{(1)})
+  \alpha_s^2 \,  C_F^2 \left(I^{(2)}_{n,C_F^2} -I_n^{(1)}\cdot I_s^{(1)} +\frac{1}{2}[I_s^{(1)}]^2\right) \nn \\
&&+ \, \alpha_s^2 \,  C_F C_A \left(I^{(2)}_{n,C_F C_A} - I^{(2)}_{s,C_F C_A}\right) +O(\alpha_s^3) \, .
\eea
The $I^{(k)}_n$ and $I^{(k)}_{s}$ are the $O(\alpha_s^k)$ contributions to the naive
collinear and soft matrix elements, respectively. We have made powers of $\alpha_s$ and the color factors accompanying the diagrams explicit. 
$I_{n,C_F^2}^{(2)}$ and $I_{n,C_F C_A}^{(2)}$ denote the contributions from two-loop collinear diagrams proportional
to $C_F^2$ and $C_F C_A$, respectively. $I_{s, C_F C_A}^{(2)}$ is the contribution from two-loop soft diagrams
that are proportional to $C_F C_A$. The two-loop soft contribution multiplying the $C_F^2$ term is given by
$[I_s^{(1)}]^2/2$ according to the exponentiation theorem. 
In Ref.~\cite{Idilbi:2007ff}, we verified that the zero-bin subtractions for the purely collinear matrix 
element reproduced the terms in Eq.~(\ref{tlexp}) proportional to  $C_F$ and $C_F^2$. 
In this section we will check the final term in Eq.~(\ref{tlexp}), which indicates that the two-loop terms 
proportional to $C_F C_A$ in the zero-bin subtraction  and the soft
function must be equal. 

The two-loop zero-bin contribution to the naive collinear 
matrix element contains diagrams where all momenta are soft and also mixed collinear-soft 
zero-bins where one momentum is collinear and the other is soft. Therefore, Eq.~(\ref{tlexp})
implies a nontrivial cancellation  among Feynman diagrams with mixed collinear-soft momenta. We will verify this
cancellation below. For the remainder of this section we will drop $C_F C_A$ from the subscript on $I^{(2)}_n$ and $I^{(2)}_s$ since 
this is the only contribution we are concerned with.

The purely collinear contribution to a two-loop integral with integrand ${\cal I}(k,l)$ is given 
by~\cite{Idilbi:2007ff}
\bea\label{mastzb}
\int_{k,l} \left( {\cal I}(k^c,l^c) - \bigg[ {\cal I}(k^c,l^s) -{\cal I}_{L_3}(k^s,l^s) \bigg]
-   \bigg[{\cal I}(k^s,l^c) -{\cal I}_{L_2}(k^s,l^s)  \bigg] - {\cal I}_{L_1}(k^s,l^s) \right) \, .
\eea
Here $\int_{k,l} \equiv \int \frac{d^D k}{(2\pi)^D} \frac{d^D l}{(2\pi)^D}$, and we use the notation of Ref.~\cite{Idilbi:2007ff}. 
The term $\int_{k,l} {\cal I}(k^c,l^c)$ is the naive collinear contribution. For this contribution,
the integrand is evaluated assuming collinear scaling for both momenta, which is denoted by the 
superscript $c$ on $k$ and $l$ in Eq.~(\ref{mastzb}).
The remaining terms in Eq.~(\ref{mastzb}) are the zero-bin contributions which must be subtracted.
There is the zero-bin arising when both $k$ and $l$ are soft, denoted by
\bea
\int_{k,l} {\cal I}_{L_1}(k^s,l^s) \, .
\eea
The superscript $s$ denotes that the momenta are taken to be soft, and $L_1$ indicates that 
we are taking $k$ and $l$ to the soft region simultaneously~\cite{Idilbi:2007ff}.
There are also mixed collinear-soft zero-bins
that arise when either $k$ is collinear and $l$ is soft,
\bea
\int_{k,l}\bigg[ {\cal I}(k^c,l^s) -{\cal I}_{L_3}(k^s,l^s) \bigg] \, ,
\eea
or when  $l$ is collinear and $k$ is soft,
\bea
\int_{k,l}\bigg[{\cal I}(k^s,l^c) -{\cal I}_{L_2}(k^s,l^s)  \bigg]  \, .
\eea
The $L_2$ limit is defined by taking $k$ soft while $l$ is collinear, then taking $l$ to be soft.
The $L_3$ limit is the same with $k$ and $l$ interchanged.
To see that the  limits are in general different, consider what happens to 
a propagator with momentum $k+l$ in the three limits:
\bea\label{llimits}
\underset{L_1}{{\rm lim}} \,\frac{1}{(k+l)^2} &=&  \frac{1}{(k+l)^2} \, , \nn \\
\underset{L_2}{{\rm lim}} \,\frac{1}{(k+l)^2} &=& \frac{1}{l^2+ \nb \cdot l \, n \cdot k} \, , \nn \\
\underset{L_3}{{\rm lim}} \,\frac{1}{(k+l)^2} &=& \frac{1}{k^2+ \nb \cdot k \, n \cdot l} \, .
\eea
When $k$ and $l$ are collinear, $k^2$, $2 k\cdot l$, and $l^2$ are all $O(\lambda^2)$.
Taking $k$ and $l$ to the soft region simultaneously, $k^2$, $2 k\cdot l$, and $l^2$ are 
all $O(\lambda^4)$. The propagator in Eq.~(\ref{llimits}) is $O(\lambda^{-4})$ in the $L_1$ limit, 
as opposed to $O(\lambda^{-2})$ when $k$ and $l$ are collinear. However, the relative importance 
of all three terms in the denominator remains the same, so the form of the propagator
is unchanged.
In the $L_2$ limit, we first take $k$ soft while keeping $l$ collinear. Then $l^2$ and
$\nb \cdot l\,  n\cdot k$ are $O(\lambda^2)$ but  $k^\perp \cdot l^\perp$ is $O(\lambda^3)$
and $n\cdot l \,\nb \cdot k$ and $k^2$ are $O(\lambda^4)$. The $O(\lambda^3,\lambda^4)$ terms
in the denominator are dropped. Next, we take $l$ soft. The propagator denominator is 
still $l^2 +\nb\cdot l \, n \cdot k$, but now $l^2$ and $\nb\cdot l \, n \cdot k$ are 
$O(\lambda^4)$. The propagator scales as $O(\lambda^{-4})$ in the $L_1$, $L_2$, and $L_3$ 
limits, but the form  of the propagator is different in each case.

Strictly speaking, Eq.~(\ref{mastzb}) is naive since it does not account for the possibility 
that a mixed collinear-soft zero-bin arises when one linear combination of $k$ and $l$ becomes soft
and the orthogonal linear combination stays collinear. One must check for a zero-bin for every linear
combination of loop momenta that appears in a propagator of the Feynman diagram. However, in practice
we find that the zero-bin contribution is subleading in $\lambda$ unless the gluon propagator
connects with the Wilson line. Therefore if we route the momenta so that the momenta of each 
gluon connected to the Wilson line coincides with one of the loop momenta, then 
Eq.~(\ref{mastzb}) is sufficient.

\begin{figure}[t]
 \begin{center}
 \includegraphics[width=5.0in]{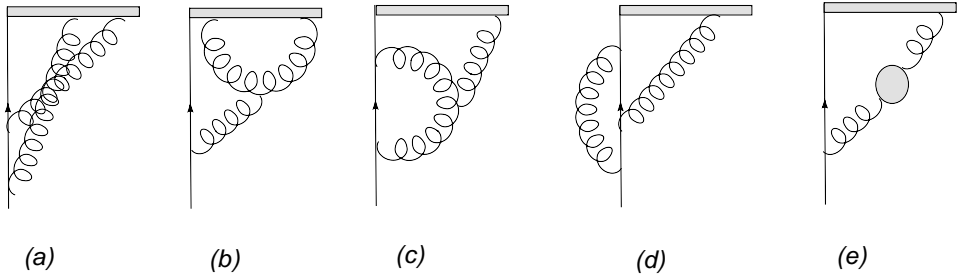}
 \end{center}
 \vskip -0.7cm \caption{Two-loop SCET diagrams contributing to $I^{(2)}_{n,C_F C_A}$.
 The solid line with an arrow is the incoming massless quark. The double line is the Wilson line.
 All gluons are collinear.}
  \label{nona}
 \end{figure}

\begin{figure}[t]
 \begin{center}
 \includegraphics[width=5.0in]{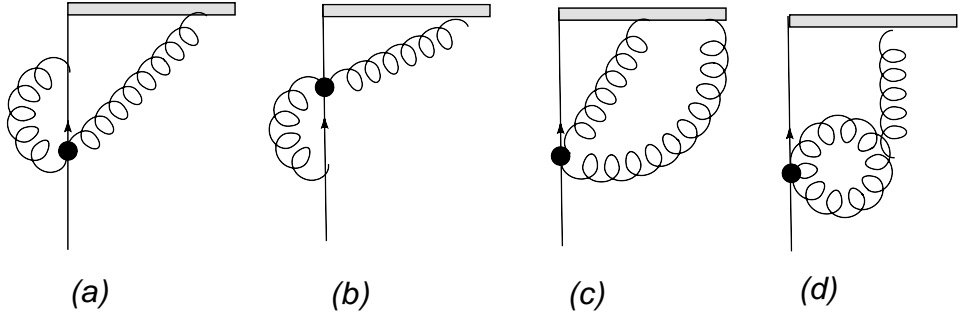}
 \end{center}
 \vskip -0.7cm \caption{Two-loop SCET diagrams involving the seagull vertex.}
  \label{seagull}
 \end{figure}

\begin{figure}[!t]
 \begin{center}
 \includegraphics[width=5.0in]{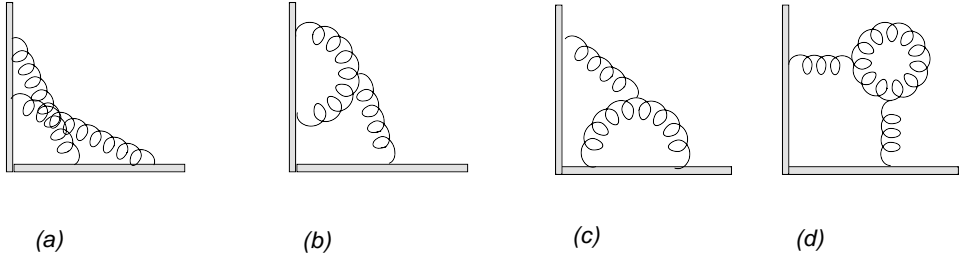}
 \end{center}
 \vskip -0.7cm \caption{Two-loop diagrams contributing to $I_s^{(2)}$.}
\label{nonasoft}
\end{figure}

Next we turn to the two-loop collinear SCET diagrams that give a contribution proportional to $C_F C_A$.
Two-loop QCD-like graphs are depicted
in Fig.~\ref{nona}, while two-loop graphs with the SCET seagull vertex are shown in Fig.~\ref{seagull}.
We will analyze diagrams in Fig.~\ref{nona} first, and start by considering the zero-bin from the $L_1$ limit. 
It is easy to see that this contribution is identical to the contribution of the two-loop soft matrix element
proportional to $C_F C_A$. All calculations are performed in Feynman gauge. The external quark is on-shell and massless.
Because the equivalence relies on field redefinitions, we cannot regulate IR divergences by taking 
external particles off-shell. We will use DR to regulate both UV (ultraviolet) and IR divergences.
We denote the momentum of the outermost gluon connected 
to the Wilson line as $k$. In Figs.~\ref{nona}(a) and (b), the inner gluon connected to the Wilson
line has momentum $l$. In Figs.~\ref{nona}(c) and (d), $l$ is the momentum of one of the 
gluons in the quark-gluon vertex correction subgraph. In Fig.~\ref{nona}(e), the blob represents all 
possible contributions to the gluon self-energy that give rise to a color factor $C_FC_A$. The loop momentum $l$ 
is the momentum of one of the gluons or ghosts in the  self-energy subgraph. Then the propagator denominators of all gluons (or ghosts) in Fig.~\ref{nona} are $l^2$, $k^2$, or $(l+k)^2$. From Eq.~(\ref{llimits}) we see that these propagators are unchanged in the $L_1$ limit. The coupling of the gluons  to the Wilson line is also unaffected by taking the $L_1$ limit. Finally, we need to consider the modification to the coupling of the collinear quark to the soft gluon.
In the $L_1$ limit, expanding to lowest order in $\lambda$, it is easy to show that the 
collinear quark coupling and propagator give the same Feynman rules as the eikonal Wilson line, $U_n^\dagger$.
No analog of Fig.~\ref{nona}(d) exists in the soft matrix element. In the $L_1$ limit
of the collinear matrix element, Fig.~\ref{nona}(d) gives a contribution that is subleading in 
$\lambda$.  The $L_1$ limit
of the other diagrams in Fig.~\ref{nona} are exactly the same as the corresponding 
two-loop diagrams contributing to the $C_F C_A$ term in the calculation of the 
soft matrix element, shown in Fig.~\ref{nonasoft}.
It is then obvious that 
\bea
\int_{k,l} {\cal I}^{(2,a-e)}_{L_1}(k^s,l^s) = I^{(2)}_{s} \, ,
\eea
where ${\cal I}^{(2,a-e)}(k,l)$ denotes the sum of the integrands coming from the
two-loop diagrams in Fig.~\ref{nona}(a)-(e).
Thus the zero-bin subtraction and soft Wilson line subtraction give equivalent results if the remaining 
zero-bin contributions all vanish.
The remaining possible zero-bin contributions are the mixed collinear-soft
zero-bins of the diagrams in Fig.~\ref{nona} and zero-bin contributions
from diagrams with the SCET seagull vertex depicted in Fig.~\ref{seagull}. 
Note that the mixed collinear-soft zero-bin subtraction does not vanish for 
the two-loop Abelian diagrams and is necessary to reproduce the 
$-\alpha_s^2 C_F^2 I_n^{(1)}\cdot I_s^{(1)}$ term in Eq.~(\ref{tlexp}).

We next analyze the mixed collinear-soft zero-bin contributions from Fig.~\ref{nona}.
With the routing described above it is straightforward to show that the zero-bin subtraction
is subleading in $\lambda$ unless it comes from the region where $k$ is soft and $l$ is collinear.
There is no mixed collinear-soft zero-bin contribution from Fig.~\ref{nona}(e).
The reason is that when $l$ is collinear and $k$ soft, 
Fig.~\ref{nona}(e) 
gives a contribution proportional to 
\bea\label{enaive}
\nb^{\mu}n^{\nu}\int_{k,l}\frac{k^2g^{\mu\nu} - k^\mu k^\nu}
{k^2 \, \nb\cdot k \, n\cdot k\,l^2\, (l^2+\nb\cdot l \,n\cdot k)}
\eea
From the naive collinear integrand in Eq.~(\ref{enaive}) one must subtract
the integrand in the $L_2$ limit. However, the integrand in Eq.~(\ref{enaive}) does not 
change in the $L_2$ limit, so the integrand when $l$ is  collinear and $k$ is soft 
vanishes,
\bea
{\cal I}^{(2,e)}(k^s,l^c)-{\cal I}^{(2,e)}_{L_2}(k^s,l^s)=0 \, .
\eea

The remaining diagrams, Figs.~\ref{nona}(a)-(d), have nonvanishing mixed collinear-soft zero-bin
contributions. There is a nontrivial cancellation among the four diagrams that ensures that the
two-loop zero-bin subtraction proportional to $C_F C_A$ reproduces the right hand side of Eq.(\ref{tlexp}).
We will describe the evaluation of Fig.~\ref{nona}(b) in some detail and quote our results for the 
remaining diagrams. Evaluation of Fig.~\ref{nona}(b) gives 
\bea
g_s^4 C_A C_F \int_{k,l} \frac{\nb \cdot (l-k) \, \nb\cdot( p+l+k)}{l^2 \, k^2 \,(l+k)^2 \, \nb \cdot (l+k) \, \nb \cdot k \,(p+l+k)^2}\,.
\eea
When $k$ and $l$ are taken  to be soft and collinear, respectively, we get the naive collinear
contributions to $I^{(2)}$,
\bea\label{ncs}
I^{(2,b)}_{n, \rm nc} &\equiv& \int_{k,l} {\cal I}^{(2,b)}(k^s,l^c) \nn \\
&=&(4\pi)^2 \int_{k,l} \frac{1}{k^2 \, \nb \cdot k}
 \frac{\nb \cdot (p+l) }{l^2 \, (l^2 + \nb\cdot l\, n\cdot k) \,((p+l)^2+ \nb\cdot(p+l) \,n \cdot k)} \,.
\eea
The factor of $(4\pi)^2$ arises because  $I^{(2,b)}_{n, \rm nc}$ is defined to be the amplitude of the Feynman diagram
divided by $\alpha_s^2 C_F C_A$.
The $l$-integral is easily evaluated using Feynman parameters and the result is
\bea
I^{(2,b)}_{\rm nc}&=&I_s^{(1)} \times
\frac{1}{\e_{\rm IR}^2} \frac{\Gamma[1-\eir]\Gamma[2-\eir]\Gamma[1+\eir]}{\Gamma[2-2\eir]} \, ,
\eea
where 
\bea
I_s^{(1)} \equiv
i  \int_k \frac{1}{k^2\, \nb\cdot k \, n\cdot k}
\left(\frac{\nb \cdot p \,n \cdot k}{4\pi \mu^2}\right)^{-\e}\, .
\eea
The divergences are of IR origin and this is denoted by the subscript ${\rm IR}$ 
on the DR parameter $\e = (4-D)/2$. 

Note that the subgraph containing the virtual $l$ momentum gives rise
to a  double IR pole, $1/\e_{\rm IR}^2$. In a one-loop diagram, double  $1/\e_{\rm IR}^2$ poles come from regions
where the virtual momenta is both soft and collinear. If $l$ is purely collinear we do not 
expect to see a double IR divergence in this subgraph. This arises in the naive collinear
graph because we have not excluded the region where $l$ is soft. Once we perform the zero-bin
subtraction (by removing the $L_2$ region) the final answer must be free from $1/\e_{\rm IR}^2$
poles. We will see that this is the case below.

Taking the $L_2$ limit of the integrand in Eq.~(\ref{ncs}) yields
\bea\label{tlzb}
I^{(2,b)}_{n,L_2}\equiv \int_{k,l} {\cal I}^{(2,b)}_{L_2}(k^s,l^s) 
&=&(4\pi)^2 \int_{k,l} \frac{1}{k^2 \, \nb \cdot k} \frac{1}{l^2 \, (l^2 + \nb\cdot l\, n\cdot k) \,n \cdot(l+ k)} \nn \\
&=&-(4\pi)^2 \int_{k,l} \frac{1}{k^2 \, \nb \cdot k} \frac{1}{l^2 \, (l^2 + \nb\cdot l\, n\cdot k) \,n \cdot l} \, ,
\eea
where the last line follows from the change of variables $\nb \cdot l \to - \nb \cdot l$
and $n \cdot l \to -n\cdot l-n \cdot k$. The same integral arises in the evaluation of the 
$L_2$ limit of Fig.~\ref{nona}(c). 
The evaluation of this integral is subtle and is discussed in the Appendix. The result is 
\bea
I_{L_2}^{(2,b)}&=&-I_s^{(1)} \times
\left(\frac{1}{\euv} - \frac{1}{\eir}\right) \frac{1}{\eir} \, ,
\eea
so the mixed collinear-soft zero-bin contribution from Fig.~\ref{nona}(b) is
\bea\label{bpc}
I^{(2,b)}_{\rm pc}&=& I^{(2,b)}_{\rm nc}-I^{(2,b)}_{L_2} \nn \\
&=& I_s^{(1)} \times
\left[\frac{1}{\e_{\rm IR}^2} \frac{\Gamma[1-\eir]\Gamma[2-\eir]\Gamma[1+\eir]}{\Gamma[2-2\eir]} 
+\left(\frac{1}{\euv}-\frac{1}{\eir}\right)\frac{1}{\eir}\right] \, .
\eea
As expected the $1/\eir^2$ poles cancel in the purely collinear loop integral.

We obtain a similar result for Fig.~\ref{nona}(a), 
\bea\label{apc}
I^{(2,a)}_{\rm pc}&=&I_s^{(1)} \times \left[
-\frac{2}{\e_{\rm IR}^2} \frac{\Gamma[1-\eir]\Gamma[2-\eir]\Gamma[1+\eir]}{\Gamma[2-2\eir]}
-2\left(\frac{1}{\euv}-\frac{1}{\eir}\right)\frac{1}{\eir}\right] \, .
\eea
When evaluating Fig.~\ref{nona}(c) in the limit that $k$ is soft and $l$ is collinear, we find that
the $l$-integral can be split into a UV divergent term, an IR divergent term, and a
term that is identical to the $l$-integral of Eq.~(\ref{enaive}). The last term gives a vanishing contribution
once the $L_2$ limit of the integrand is subtracted. The remaining terms give
\bea\label{cpc}
I^{(2,c)}_{\rm pc} = I_s^{(1)} &\times& \left[-\frac{\Gamma[\euv]\Gamma[2-\euv]\Gamma[1-\euv]}{\Gamma[2-2\euv]}
+\frac{1}{\e_{\rm IR}^2} \frac{\Gamma[1-\eir]\Gamma[2-\eir]\Gamma[1+\eir]}{\Gamma[2-2\eir]} \right. \nn \\
&& \left. \quad
+\left(\frac{1}{\euv}-\frac{1}{\eir}\right)\frac{1}{\eir}\right] \, .
\eea
Finally, Fig.~\ref{nona}(d) gives
\bea\label{dpc}
I^{(2,d)}_{\rm pc}&=&I_s^{(1)} \times \frac{\Gamma[\euv]\Gamma[2-\euv]\Gamma[1-\euv]}{\Gamma[2-2\euv]}\, .
\eea
We see that 
\bea
I_{\rm pc}^{(2,a)} + I_{\rm pc}^{(2,b)} +I_{\rm pc}^{(2,c)} +I_{\rm pc}^{(2,d)} = 0 \, ,
\eea
so the mixed collinear-soft zero-bin contributions from Figs.~\ref{nona}(a)-(e) add up to zero.

In addition to the mixed collinear-soft zero-bin discussed above there are two-loop SCET diagrams 
that have no QCD analog. These graphs, given in Fig.~\ref{seagull}, involve the seagull interactions with two collinear quarks
and two collinear gluons (see  Fig.~1 of Ref.~\cite{Bauer:2000yr}). The Feynman rule for the SCET seagull
interaction is such that if a collinear gluon from the seagull vertex is contracted
with the gluon coming from the Wilson line, $W_n^{\dagger}$, the graph vanishes, so 
the diagrams of Fig.~\ref{seagull}(a)-(c) are zero.  For Fig.~\ref{seagull}(d) we obtain 
\bea
-\frac{1}{2}g_s^4 C_F C_A\int_{k,l}\frac{\nb\cdot(p-k) \, \nb\cdot(2 \,l-k)}{\nb\cdot k \, k^2 \, l^2 \, (l-k)^2
\, (p-k)^2}\left[\frac{1}{\nb\cdot(p-l)}-\frac{1}{\nb\cdot(p+l-k)}\right]\, .
\eea
When  $k$ is soft and $l$ collinear one obtains the naive collinear contribution to $I^{(2)}_n$: 
\bea
I^{(2,3d)}_{\rm nc} = (4\pi)^2 \int_{k,l}\frac{1}{\nb\cdot k \,  n\cdot k \, k^2}\frac{\nb\cdot l}{l^2 \,(l^2-\nb\cdot l n\cdot k)}\left[\frac{1}{\nb\cdot(p-l)}-\frac{1}{\nb\cdot(p+l)}\right]\, ,
\eea
which by SCET power counting is ${\cal O}(1)$. The integration over $l$ can be performed by combining denominators using the standard Feynman parameterization and after completing 
the square  the resulting integral vanishes by symmetry. Therefore this zero-bin contribution vanishes. It is also easy to check that the zero-bin contributions from both the $L_1$ and 
$L_2$ regions are subleading in the $\lambda$ expansion and can be ignored.
This result is essential as Fig.~\ref{seagull}(d) has no analog in the soft function diagrams shown in Fig.~\ref{nonasoft}.

\section{Conclusion}

In this paper, we completed the two-loop analysis of zero-bin subtractions to the jet function appearing in the quark form factor
initiated in Ref.~\cite{Idilbi:2007ff}. We verified that the zero-bin subtraction is equivalent to dividing by a matrix element 
of soft Wilson lines to two-loop order. We also supplied a proof of the factorization of the naive collinear Wilson line into a purely collinear 
Wilson line and a soft Wilson line, valid to $O(\lambda^2)$.  This property is essential for the argument for the  equivalence 
of soft and zero-bin subtractions first presented in  Ref.~\cite{Lee:2006nr}. These arguments imply that the equivalence of soft and zero-bin
subtractions should hold to all orders in perturbation theory.

One important consequence of this  equivalence (to lowest order in $\lambda$) is that it provides an operator definition for the zero-bin 
subtraction procedure that goes beyond perturbation theory. The equivalence may also help to simplify higher order calculations in SCET. For example, in the calculation of the jet function in this paper, we saw that the equivalence led us to anticipate a cancellation between
mixed collinear-soft zero-bin subtractions coming from several different SCET diagrams. We expect that the equivalence 
will simplify SCET calculations of collinear correlation functions, like parton distribution functions, jet functions, and 
fragmentation functions, since it allows one to work with the naive collinear correlation functions and soft matrix elements,
rather than having to remove zero-bins in collinear correlation functions diagram by diagram. 

Though the analysis of this paper, as well as Refs.~\cite{Lee:2006nr} and \cite{Idilbi:2007ff}, focuses on avoiding double counting in factorization theorems for inclusive processes, the issue of double counting arises in exclusive processes as well~\cite{Manohar:2006nz}. It would be interesting to see if one could find an operator definition of the 
zero-bin subtraction in SCET$_{\rm II}$.

\acknowledgments 
 This work was supported in part by the Department
of Energy under grant numbers DE-FG02-05ER41368, DE-FG02-05ER41376, and DE-AC05-84ER40150.
We thank S.~Fleming for useful comments.
 
\section{Appendix}

In this Appendix, we discuss our evaluation of the  $l$ integral in Eq.~(\ref{tlzb}). First, we use the identity
\bea\label{lint}
\frac{1}{l^2 \, (l^2 +\nb \cdot l \, n\cdot k) n\cdot l} 
= 2 \int_0^\infty d\lambda \int_0^1 dx  \frac{x}{(l^2 +x \,\lambda \, n\cdot l +(1-x) \,\nb \cdot l\, n\cdot k)^3}
\,.\eea
The parameter $\lambda$ is dimensionful and can be made dimensionless by rescaling 
$\lambda \to \nb \cdot p \, \lambda$. The $l$ integral is then straightforward to evaluate
by completing the square. The result is 
\bea\label{ans1}
I_{L_2}^{(2,b)}&=&I_s^{(1)} \times
\int_0^\infty d\lambda \, \lambda^{-1-\e}\int_0^1dx \, x^{-\e}(1-x)^{-1-\e} \,\Gamma[1+\e]
 \nn \\
&=&-I_s^{(1)}\times \left(\frac{1}{\euv} - \frac{1}{\eir}\right) 
\frac{1}{\eir}\,.
\eea
The $\lambda$ integral is proportional to $1/\e_{\rm UV} -1/\e_{\rm IR}$ while the 
$1/\e$ from the $x$ integral is clearly IR in origin. Note that we have set a factor
of 
\bea
\frac{\Gamma[1-\e]^2 \Gamma[1+\e]}{\Gamma[1-2\e]}\, ,
\eea
equal to 1. Anything  besides the double $1/\e$ poles in the evaluation of the 
scaleless integral is ambiguous. One source of ambiguity is the freedom to rescale 
$\lambda$. We chose to rescale $\lambda$ so as to obtain a prefactor proportional
to $(\nb \cdot p\, n \cdot k/(4\pi \mu^2))^{-\e}$ , but other rescalings are possible.
The other source of ambiguity is due to the ambiguity in expanding any function
of $\e$ when multiplying a factor of $1/\euv-1/\eir$.

However, there is even more ambiguity in the result because the coefficient of the double 
$\e$ poles actually depends on how one choses to combine denominators using Feynman parameters.
To see this we now evaluate Eq.~(\ref{tlzb}) using the  identity
\bea
\frac{1}{l^2 \, (l^2 +\nb \cdot l \, n\cdot k) n\cdot l} 
= 2 \int_0^\infty d\lambda \int_0^1 dx  \frac{x}{(l^2 +x \,\lambda \, n\cdot l + x \,\nb \cdot l\, n\cdot k)^3}\, .
\eea
The result is then
\bea\label{ans2}
I_{L_2}^{(2,b)}&=&I_s^{(1)} \times
\int_0^\infty d\lambda \, \lambda^{-1-\e}\int_0^1dx \, x^{-1-2\e} \,\Gamma[1+\e]
 \nn \\
&=&-\frac{1}{2}I_s^{(1)}\times \left(\frac{1}{\euv} - \frac{1}{\eir}\right) 
\frac{1}{\eir}\, ,
\eea
which differs from Eq.~(\ref{ans1}) by a factor of 1/2!

Mathematically, there is no inconsistency here, since $\euv = \eir = 2-D/2$
so the right hand sides of Eq.~(\ref{ans1}) and Eq.~(\ref{ans2}) are both equal to 
zero. Both results are consistent with the well-known result that scaleless integrals
vanish in DR. However, if a physical regulator is used, a scaleless  integral would be the sum
of UV and IR divergent terms. Since the cancellation of UV and IR divergences in physical
observables is handled differently 
in quantum field theory, one often wishes to separate scaleless integrals into their UV and IR divergent 
parts, even if DR is being used to regulate both. Evidently, for the double $1/\e$ poles appearing in 
the integral of Eq.~(\ref{tlzb}), such a separation is ambiguous. 

Faced with this situation, one is forced to invoke a prescription for handling the zero-bin integrals in the
$L_2$ limit. One physically motivated prescription was mentioned earlier, that is to fix the 
overall coefficient so that the $1/\eir^2$ poles cancel in the purely collinear integral.
This leads to the result of Eq.~(\ref{ans1}). 
Another way to resolve the ambiguity is to modify the integral by including a $\lambda$-suppressed correction. 
For instance, we can replace $l^2  + \nb \cdot l \, n \cdot k$
with $(l+k)^2$ with $k^2\neq 0$. Now the result of the integral is completely unambiguous
and equal to the result of Eq.~(\ref{ans1}) with the following modification
\bea
\left(\frac{1}{\euv}-\frac{1}{\eir}\right) \frac{1}{\eir}
\to \left(\frac{1}{\euv}-\log\left(\frac{-k^2}{4\pi \mu^2}\right) + \frac{\pi^2}{12}\right) \frac{1}{\eir}\,.
\eea
This prescription unambiguously fixes the coefficient of the mixed $1/(\euv \eir)$ pole and is
in agreement with the result of Eq.~(\ref{ans1}). We will use the result of Eq.~(\ref{ans1}) for
the evaluation of the $l$-integral for the $L_2$ zero-bin integral in Eq.~(\ref{tlzb}).

\end{document}